%% file: main.tex
\documentclass[twocolumn]{aastex7}

\usepackage{xspace}

\newcommand{\eg}{e.g.\xspace}

\newcommand{\kms}{km~s\ensuremath{^{-1}}\xspace}

\newcommand{\Nifs}{\ensuremath{^{56}}Ni\xspace}

\newcommand{\jwst}{{\it JWST}\xspace}

\newcommand{\HeI}{He\hspace{0.25em}{\sc i}}

\newcommand{\OI}{O\hspace{0.25em}{\sc i}}

\newcommand{\CI}{C\hspace{0.25em}{\sc i}}

\newcommand{\NaI}{Na\hspace{0.25em}{\sc i}}

\newcommand{\MgI}{Mg\hspace{0.25em}{\sc i}}
\newcommand{\ArII}{Ar\hspace{0.25em}{\sc ii}}

\newcommand{\FeII}{Fe\hspace{0.25em}{\sc ii}}

\newcommand{\CoII}{Co\hspace{0.25em}{\sc ii}}

\newcommand{\NiII}{Ni\hspace{0.25em}{\sc ii}}
\newcommand{\NiIII}{Ni\hspace{0.25em}{\sc iii}}

\newcommand{\E}{$E\left(B-V\right)$}

\newcommand{\R}{$f_\mathrm{em}/f_\mathrm{abs}$}

\newcommand{\ggi}{SN~2024ggi}

\begin{document}

\title{JWST Observations of SN 2024ggi II: NIRSpec Spectroscopy and CO Modeling at +285–385 Days Past the Explosion}

\input authors

\submitjournal{ApJ}

\received{\today}
\revised{\today}
\accepted{\today}

\begin{abstract}

We present \textit{James Webb Space Telescope} (JWST) NIRSpec observations of SN~2024ggi, spanning wavelengths of 1.7--5.5\,$\micron$ at +285.51 and +385.27 days post-explosion. These nebular spectra are dominated by asymmetric emission lines from atomic species including H, Ca, Ar, C, Mg, Ni, Co, and Fe, indicative of an aspherical explosion. The other strong features are molecular CO vibrational bands from the fundamental and first overtone.
We introduce a novel, data-driven approach using non-LTE 3D radiative transfer simulations to model the CO emission with high fidelity. This method enables us to constrain the three-dimensional CO distribution and its radial temperature structure. CO formation is found to occur prior to day +285, with subsequent evolution characterized by progressive evaporation. The CO mass decreases from approximately (7.9 to 1.0)$\times{\rm 10^{-4}~ M_\odot}$, while the average temperature drops from $\approx $ 2300 K to $\approx $ 1800 K. Concurrently, the CO distribution becomes increasingly clumped ($f_c\approx$ 5.5 to 10). The minimum velocity of the CO-emitting region decreases ($v_1\approx$  1400 to 800 \kms), above the receding photosphere velocity (${\rm v_{ph} \approx 500~ \kms}$), suggesting the photosphere resides within Si-rich layers. However, the temperature profile indicates that only a narrow zone reaches the conditions necessary for SiO formation. Due to a lack of observational constraints, SiO clumping is not modeled, and thus, synthetic SiO profiles for mass estimates are not highlighted. We discuss the implications of these findings for dust formation processes in SN~2024ggi.
\end{abstract}

\keywords{\uat{Supernovae}{SN2024ggi}, \uat{JWST}{}, \uat{radiation transfer}{}, \uat{molecular processes}{}}

\section{Introduction} 
The origin of cosmic dust in the early universe remains a central question in astrophysics, with supernovae (SNe) frequently cited as key contributors \citep[\eg][]{Dwek_1980, Wooden_1993, Schneider_2004, Dwek_2006, Gall_2011}. Although theoretical and observational studies support their role, significant uncertainties persist regarding the quantity, composition, and grain size distribution of dust produced by various types of SNe. At the heart of this process lies molecular formation, which precedes dust condensation and plays a critical role in cooling the expanding ejecta \citep{Liu_Dalgarno_1995, Liljegren_2020}. These molecules not only provide nucleation sites for dust grains, but also regulate the thermal evolution necessary for the gas-to-dust transition \citep{Sluder_2018}.

 Obtaining multi-wavelength spectra covering the near- and mid-infrared (NIR and MIR, respectively) at wavelengths between 2.5-25 microns
of molecule-emitting regions in hydrogen-rich core-collapse supernovae (SNe~II) has historically been challenging. Nevertheless, several studies have confirmed the presence of freshly formed CO and SiO in the ejecta of SNe~II \citep{Roche_1991, Wooden_1993, Kotak_2006, Kotak_2009, Szalai_2011}. The CO first overtone at 2.2\,$\micron$ has been the most frequently detected molecular feature, owing to its accessibility with ground-based NIR spectroscopy \citep{Spyromilio_1996,  Meikle_2011, Davis_2019,Rho_2021,Medler25_HISS}. In contrast, detections of the SiO fundamental band at $\sim8.1$--$9.3$\,$\micron$ have been limited to a handful of events observed with the \textit{Spitzer Space Telescope} \citep[SST,][]{Kotak_2005, Szalai_2011}. Observations of the CO fundamental band remain especially scarce; aside from SN\,1987A \citep{Wooden_1993}, most detections are incomplete or only marginally constrained due to limited MIR spectral coverage \citep{Kotak_2005, Szalai_2011}.
The \textit{James Webb Space Telescope} (\jwst) has  begun to provide panchromatic observations of SNe II, spanning 0.4--25\,$\micron$ \citep{Shahbandeh_2024,Medler_25b,DerKacy:2025a,Baron2025_ggi}. This unprecedented wavelength coverage rapidly transforms the study of molecule and dust formation by allowing the regions of the first overtone and fundamental vibrational bands of CO and SiO to be simultaneously observed at high sensitivity and spectral resolution.

\ggi, classified as a SN~IIP with flash ionization features \citep[][]{Hoogendam_etal_2024,Zhai_etal_2024}, was discovered on 2024 April 11 (MJD~60411.14, \citealt{24ggi_atlas}) in the host galaxy NGC~3621, located at a distance of $7.24 \pm 0.20$~Mpc \citep{Saha_2006}. Its proximity makes it an ideal candidate for monitoring and tracking the evolution of molecule formation in SN~II explosions.

Here we present \jwst\ observations of SN~2024ggi from our \jwst\ Director’s Discretionary (DD) programs DD-6616 and DD-6717 
\citep{Ashall6677A, Ashall6716A}.
This program aimed to follow the infrared spectral evolution of SN~2024ggi from the plateau to the nebular phase. 

The first epoch of observations, which consists of  1.5-14~$\micron$ spectra, observed at 55~d past explosion, was presented in the first paper of this series \citet{Baron2025_ggi}.
These observations were obtained during the plateau phase and before the formation of CO which can be expected when the photosphere enters the C/O-rich core \citep{Spyromilio_1988,hoeflich_1988,Petuchowski_1989,Liu_Dalgarno_1992,Gerardy_2000,gerardy02,Kotak2005,2023shahbandeh_dust}.
Here we concentrate on the next two epochs obtained at +285.64 and +385.55~days past the explosion. The purpose of these observations is to investigate the formation and evolution of CO within the ejecta.

In Section~\ref{sec:datared}, we present the observations and describe the data reduction procedures. This is followed in Section~\ref{sec:Spec} by our analysis of the spectra, including line identifications, velocity measurements of key features, and spectral comparisons. In Section~\ref{sec:Models}, we introduce multi-dimensional non-local thermodynamic equilibrium (non-LTE) models of the CO-emitting regions. Finally, our discussion and conclusions are presented in Section~\ref{sec:Conclusions}.

\begin{figure*}
    \centering
    \includegraphics[width=0.99\linewidth]{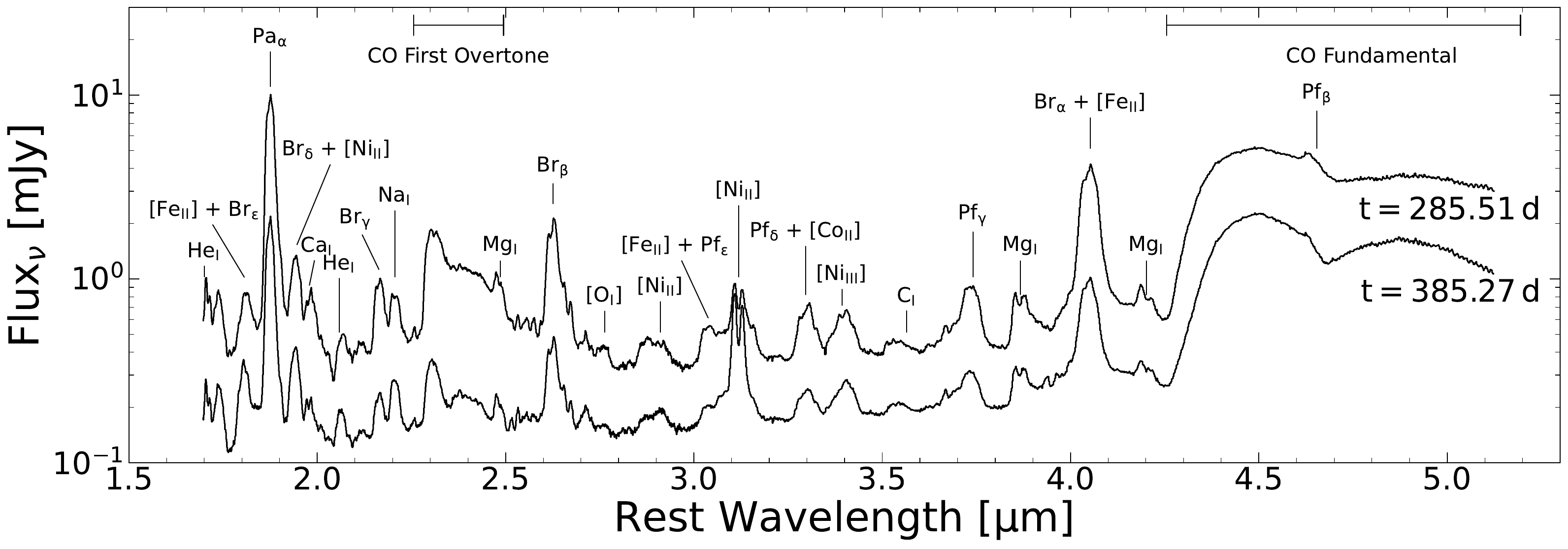}
    \caption{Our NIRSpec observations of SN~2024ggi at +285.64~d and 385.55~d past explosion. We identify the prominent spectral features.  } \label{fig:data}
\end{figure*}

\section{Data Reduction}\label{sec:datared}
We obtained two epochs of data with \jwst's Near-Infrared Spectrograph (NIRSpec; \citealt{Jakobsen_2022,Boker_2023}) as part of our program, on 2025 January 22 (MJD = 60697.13) and 2025 May 2 (MJD = 60797.29), corresponding to +285.64 and +385.55 days after the inferred explosion date of MJD = 60410.795 \citep{Jacobson24}.

For these observations, we used the F170LP/G235M and F290LP/G395M filter/grating combinations, providing spectral coverage from $1.66$--$3.07$$~\micron$ and $2.87$--$5.10$$~\micron$, respectively, at a resolving power of $R \sim 1000$. We performed the observations with the S400A1 slit, the SUBS400A1 subarray, the NRSRAPID readout pattern, 40 groups per integration and a 3-point nod dither pattern with offsets of the position Y of $+0.9845$, $0$, and $-1.1952$~arcsec.

The reduction process followed standard procedures outlined in the NIRSpec fixed-slit Jupyter notebooks \citep{law_2025}, using \texttt{jwst} pipeline version~1.18.0 \citep{2023bushouse,2025bushouse1}, Build~11.3, and the \texttt{jwst\_1364.pmap} CRDS calibration files. We adjusted standard reduction parameters to include a scaling factor in the \texttt{outlier\_detection} step during stage 3 spectroscopic processing, reducing noise spikes caused by bad pixels in the three dithers while preserving source flux. We used total exposure times of $191.695$~seconds and $472.135$~seconds for observations at +285.64~d and +385.55~d, respectively.

\section{Spectral Properties} 

\label{sec:Spec}
\subsection{Line Identification} 

Figure~\ref{fig:data} shows both epochs of our \jwst\ observations. In these phases, \ggi\ has entered the nebular stage, with spectra dominated by strong emission lines. We identify the spectral features in \ggi\ using previously published line lists from \citet{Jerkstrand_2012, Davis_2019}; and \citet{Shahbandeh_2022}. The spectra exhibit numerous hydrogen lines, including Pa$\alpha$ (1.875~$\micron$), Br$\alpha$ (4.051$~\micron$), Br$\gamma$ (2.166~$\micron$), Br$\beta$ (2.626~$\micron$), Br$\epsilon$ (1.817~$\micron$), Br$\delta$ (1.944~$\micron$), Pf$\delta$ (3.297~$\micron$), Pf$\gamma$ (3.741~$\micron$), and Pf$\beta$ (4.654~$\micron$). We also identify additional lines from ions such as \HeI\ (1.7002~$\micron$, 2.0581~$\micron$), \NaI\ (2.206~$\micron$), \MgI\ (2.486, 3.867, and 4.201~$\micron$), [\OI] (2.763~$\micron$), [\CI] (3.565~$\micron$), [\FeII] (1.809, 3.029, and 4.075~$\micron$), [\CoII] (3.286~$\micron$), [\NiII] (1.985, 2.911, 3.119~$\micron$), and [\NiIII] (3.393$~\micron$). 

Broad emission features spanning 2.2--2.4~$\micron$ and 4.5--5.17~$\micron$ trace the first overtone and fundamental vibrational bands of CO, respectively. Although the first CO overtone has been detected in many SNe \citep[\eg][]{Davis_2019, Rho_2021}, observations of the CO fundamental band remain rare due to the historical lack of spectroscopic coverage in the 2--5~$\micron$ range. To date, the only published detections of the full fundamental CO band in a core collapse SN are SN~1987A and SN~2023ixf \citep{Wooden_1993,Medler_25b}. 

Over the $\sim$100~days between our two \jwst\ epochs, the spectra evolve significantly. The hydrogen emission lines decline in flux, while the lines from intermediate-mass and iron-group elements become relatively stronger. The CO emission features also weaken over this period. Notably, the shape of the CO first overtone band evolves: higher vibrational modes fade relative to the first epoch, indicating a decrease in the temperature of the CO-emitting region (we explore this further in Section~\ref{sec:Models}). The fundamental band of CO also fades, but only by $\sim$50\% in absolute flux, in contrast to the order-of-magnitude decline observed in the first overtone. This suggests that the fundamental CO band becomes a dominant coolant in the ejecta at later times.

\subsection{Spectral Line Velocities}

\begin{figure}
    \centering
    \includegraphics[width=\linewidth]{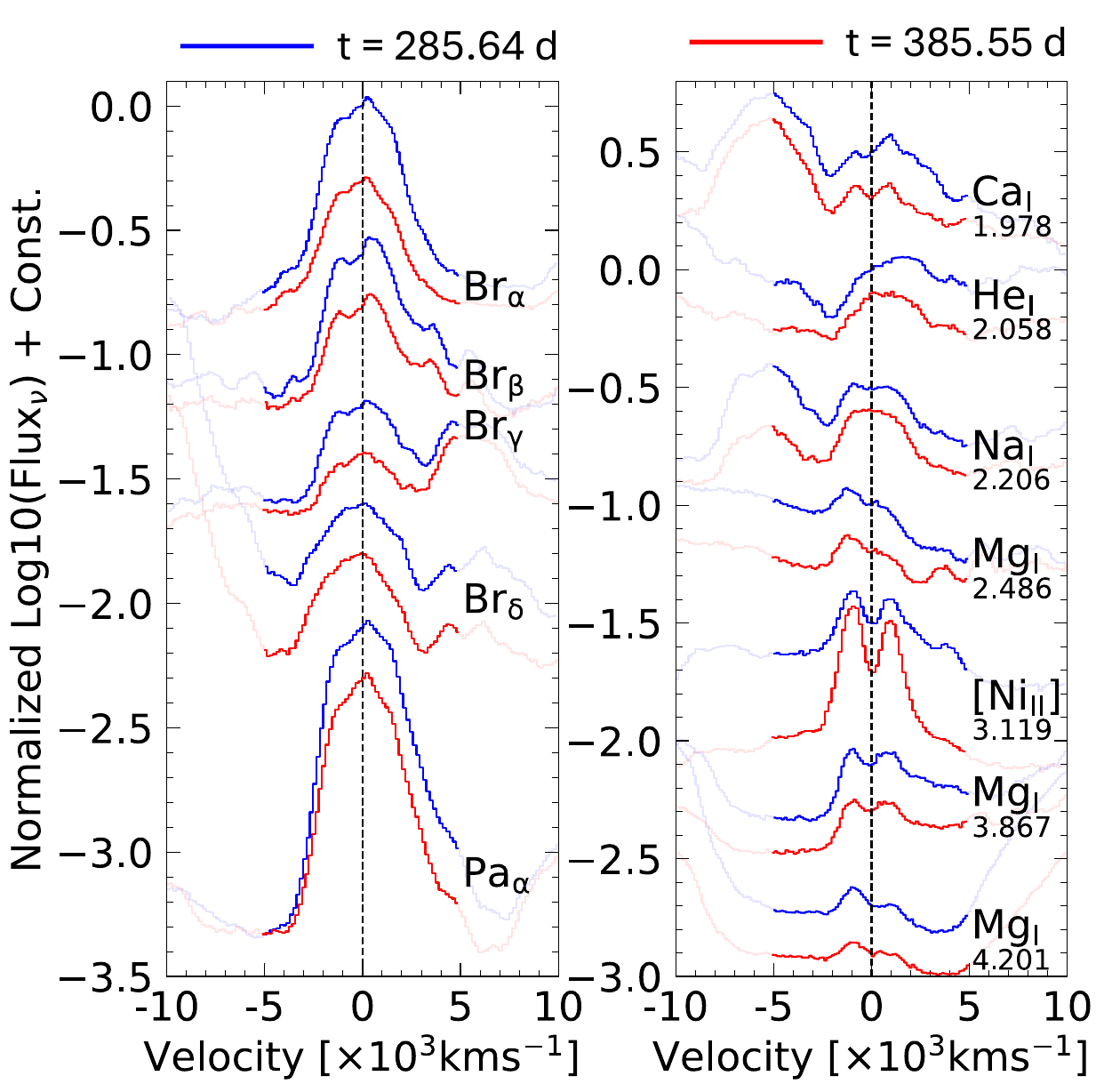}
    \caption{Spectral line profiles of the strongest ions on the nebular phase spectra of \ggi. Various hydrogen profiles are shown in the left panel, while lines of other elements are presented in the right panel.  }
    \label{fig:linevel}
\end{figure}

Figure \ref{fig:linevel} shows the spectral line velocity profiles of the most prominent features at both epochs. The hydrogen profiles span $\pm$5000~\kms\ and show no attenuation on the red side, indicating the absence of dust within the ejecta \citep{1989Lucy}. 
In fact, the Br$\alpha$, Br$\gamma$, and Pa$\alpha$ lines, which are isolated, exhibit tilted profiles with a peak on the red side at $\sim$1000~\kms, suggesting an asymmetric chemical distribution. However, the expansion velocity of these features is large, causing the profiles to blend.

Looking at the intermediate mass and iron-group elements (see the right-hand panel of Fig. \ref{fig:linevel}), we observe clear evidence of splitting in all isolated features. This is most prominent in the [\NiII]~$3.119$~$\micron$ feature, which is split by $1000  \pm  150$~\kms, as well as in the three \MgI\ features, which show slightly broader lines. The splitting of these spectral features in \ggi\ is consistent with the optical nebular phase, where similar splitting of the Fe group lines is observed \citep{2025emilio,Ferrari25}. The splitting  may be due to the ejecta being in an oblate or torus-like configuration or due to jet like structures from the explosion \citep[see e.g.][]{1999ApJ...524L.107K,   2005mazzali,2008modjaz, 2008maeda, 2011dessart}.
Overall, the splitting of the lines and attenuation of the red side of the flux in the emission features related to stellar or explosive burning, suggests that \ggi\ is an asymmetric explosion with little to no mixing between the \Nifs\ layers and the C/O-rich regions.

\subsection{Spectral Comparison} \label{sec:SpecComp}

\begin{figure}
    \centering
    \includegraphics[width=0.99\linewidth]{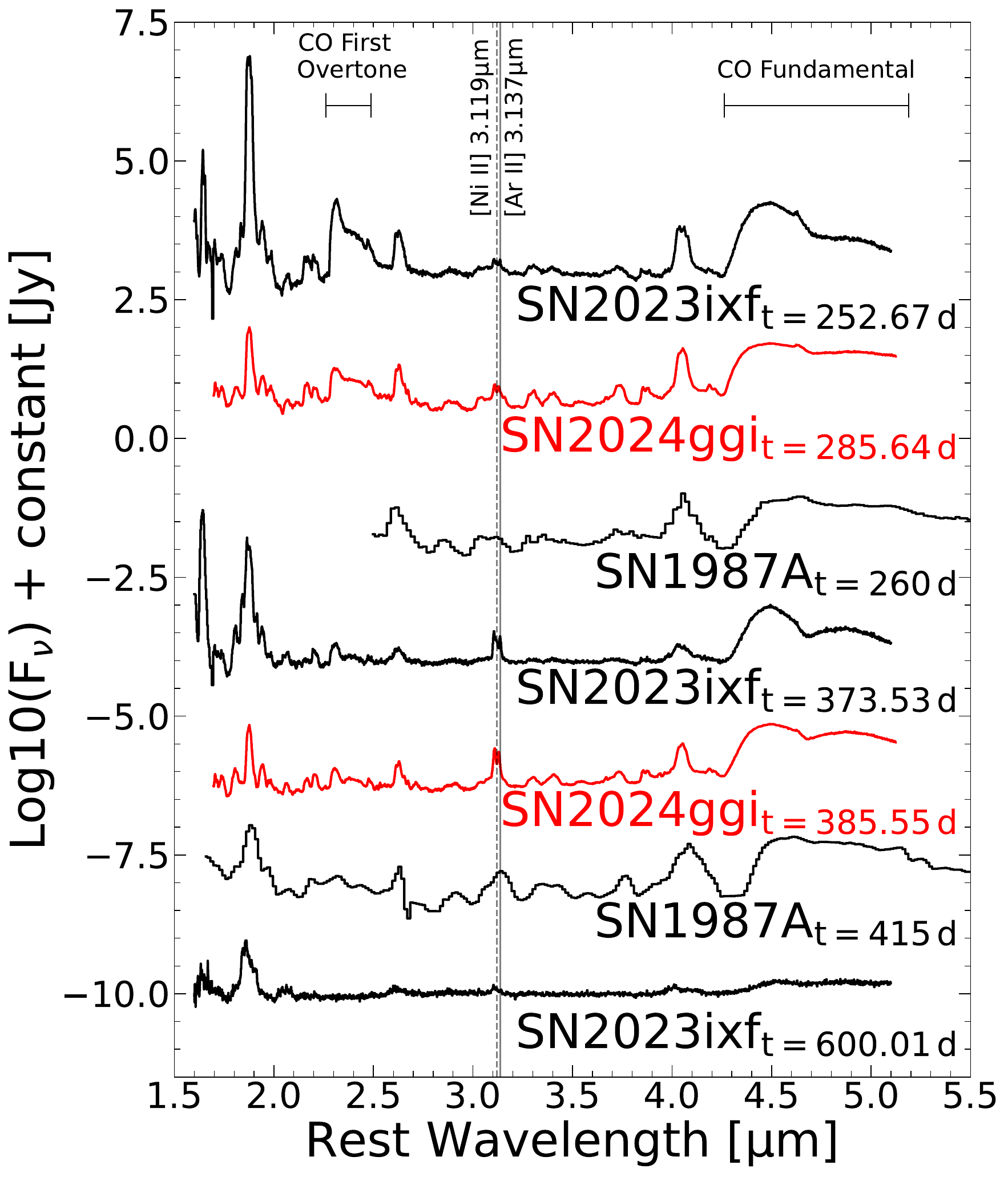}
    \caption{1.5-5.5~$\micron$ spectral comparison between \ggi, SN~1987A, and SN~2023ixf. The SN~2023ixf spectra have been continuum subtracted following the procedure of Fig. 4 in \citet{Medler_25b} to aid in visual comparison.} \label{fig:speccomp}
\end{figure}

In Fig.~\ref{fig:speccomp} we compare our \textit{JWST} spectra of \ggi\ to those of SN~1987A and SN~2023ixf, the only other SNe~II with published spectral data covering this wavelength range \citep{Wooden_1993,Medler_25b}. Overall, the spectra are remarkably similar, both dominated by emission lines from the same atomic species and by prominent CO emission. Like SN~2023ixf, \textit{JWST} data for \ggi\ were obtained at significantly higher spectral resolution, allowing us to resolve features that remained mixed in SN~1987A. For example, in \ggi\ we clearly resolve the [\NiII] and [\ArII] lines near 3.1~$\micron$, which appear to be blended in the SN~1987A data set.

The most notable difference between these objects lies in the CO first overtone region. 
In SN~1987A, the blueward extension of this feature includes CO$^{+}$ emission. This is most likely due to mixing between \Nifs and the C/O-rich layers and excitation by the emitted gamma-rays \citep{Spyromilio_1988,Petuchowski_1989,Rho_2021}.
We find no evidence of CO$^{+}$ emission in \ggi, suggesting that \Nifs\ is likely not significantly mixed in the CO-forming layers.

\section{Molecular models} \label{sec:Models}
In the following, we analyze the spectral features of CO and their evolution over time. Our goals are to: (a) quantify the abundance of diatomic molecules, (b) assess the physical conditions of the corresponding stellar layers, including their abundance, temperature, and distribution, and (c) investigate the small-scale 3D morphology of CO, and use it to constrain details on the stellar progenitor and dust formation.

Frequently, diatomic molecules are analyzed under the assumption that their features are optically thin, with a constant temperature and total mass adjusted to best fit the observed data \citep{2018Banerjee,2023stritzinger,park2025}. This treatment breaks down for the optically thick fundamental bands and at later times when non-LTE effects are important \citep{Liu_Dalgarno_1992}.  

To overcome this shortcoming, other works have employed forward modeling, combining explosion models with the time evolution of molecular formation to produce spherical non-LTE synthetic spectra (e.g., \citealp{hoeflich_1988,Liu_Dalgarno_1992,Liu_Dalgarno_1994,Kotak05,Rho_2021,McLeod24}). For some SNe, this approach provides reasonable results for the CO first overtones (see e.g. SN~1987A, \citealt{hoeflich_1988,Spyromilio_1988,Liu_Dalgarno_1992} and SN~2000ew \citealp{Gerardy_2002}), as well as for the optically thin part of the CO fundamental band (see Fig. 4 for models of SN~2005df in \citealp{Kotak05}).  However, in other SNe, such as SN~1987A, the profiles and synthetic flux levels of the CO fundamental band to the first overtone are inconsistent by factors of 2 to 5 (see Fig. 12 in \citealt{McLeod24}). This discrepancy suggests some fundamental shortcomings, including lack of understanding on how the molecules contribute to the overall temperature structure, and of how   multidimensionality effects the molecular forming layers \citep{Liljegren_2020}. Additionally, forward modeling based on spherical geometries is inherently incapable of resolving the underlying multidimensional physics. 

Thus, we developed an inverse data-driven approach to analyze the unique spectra only obtainable with \jwst.  Namely, we use the observed absolute flux, combined with full radiation transport that includes small-scale 3D morphology effects (clumping) in a homologously expanding envelope, to calculate the cooling structure by molecules and, with it, the temperature profile of the molecular layers. In turn, this can be used to
constrain the underlying physical mechanisms that drive molecular formation. This approach has been implemented in our newly-developed MOFAT code (MOlecular Fitting Analysis Tool; \citealt{Mera_2025_theory}), which allows us to analyze the observed diatomic features.

\subsection{Method for the Analysis}\label{sec:Method}

MOFAT consists of a seven parameter iterative framework that utilizes the observed flux in molecular bands, and finds the optimal solutions in the multi-dimensional parameter space. It uses physics-based modules to provide an on-demand solution for physical structures associated with the signatures of molecules formed. At its core, MOFAT assumes spherically symmetric large-scale geometry and uses a solution of the radiation transport equations that includes the effects of multi-dimensional clump-like structures. Molecular opacities are calculated by HYDRA \citep{1989HiA.....8..207S,Sharp_1990,Hoeflich_2002,Rho_2021,Hoeflich_2021_20qxp}, where the vibrational bands within each mode are treated with non-LTE correction factors \citep{Liu_Dalgarno_1992,1992gamache,2010rothman}. Unlike HYDRA, which solves the time-dependent rate equations for the formation of molecules, MOFAT takes critical parameters associated with the CO profiles and then uses observations as the driver towards an optimized solution at a particular time (see \cite{Mera_2025_theory} for a full description of MOFAT and Table \ref{tab:model_values} for a description of the parameters).

\begin{table*}
\caption{Parameter values from Figures \ref{fig:clumps} and
\ref{fig:noclumps} optimized for the two epochs observed.
\label{tab:model_values}}

{\centering
\begin{tabular}{c|c|c|c|c|c|c|c|c|c|c}
\hline
Model & Time [d] & Clumps & Mass [M$_\odot$] & T$_1$ [K] &
v$_1$ [km/s] & $\Delta v$ [km/s] & $n$ & $r/R$ & $f_c$ &
$\epsilon$ \\ \hline

P & 285 & yes & 7.997e-4 & 2298 & 1370 & 1520 & 9.50 &
1.750e-1 & 5.50 & 10.0 \\

& 385 & yes & 1.064e-4 & 1815 & 875 & 1280 & 10.15 &
1.475e-1 & 10.55 & 8.0 \\
\hline

O & 285 & yes & 7.799e-4 & 2335 & 1420 & 1500 & 9.50 &
1.800e-1 & 5.53 & 0.27 \\

& 385 & yes & 9.423e-5 & 1896 & 763 & 1080 & 10.21 &
9.510e-2 & 9.21 & 0.30 \\
\hline
\hline

S & 285 & no & 4.584e-4 & 2079 & 500 & 764 & 9.58 &
-- & -- & -- \\

& 385 & no & 8.386e-5 & 1499 & 345 & 261 & 13.99 &
-- & -- & -- \\
\hline
\hline

E1 & 285 & yes & 8.108e-4 & 2311 & 1370 & 1520 & 9.50 &
1.944e-1 & 5.50 & 10.0 \\

& 385 & yes & 1.117e-4 & 1849 & 875 & 1280 & 10.15 &
1.475e-1 & 11.36 & 8.0 \\
\hline

E2 & 285 & no & 5.225e-4 & 2079 & 700 & 764 & 9.58 &
-- & -- & -- \\

& 385 & no & 8.177e-5 & 1545 & 345 & 783 & 7.00 &
-- & -- & -- \\
\hline
\end{tabular}
\par
}

\vspace{0.5em}
\noindent
\begin{minipage}{\textwidth}
\footnotesize
\raggedright
\textbf{Note.} Models P and O
indicate the selected representative models for prolate and oblate clumps.
Models S are without clumps and provide a reasonable fit to the early epoch
(see text). To demonstrate the sensitivity on the parameters, Models E1 and
E2 are given as examples. The \textit{Clumps} column indicates whether the
model takes into account clumping effects. The \textit{Mass} column shows the
total amount of CO mass. \textit{T$_1$} shows the temperature at the inner edge
(\textit{v$_1$}) of the CO shell, which has a width of \textit{$\Delta v$}.
The tabulated \textit{n} corresponds to the inner density-profile slope $n_1$;
the outer slope $n_2$ is not shown. \textit{r/R} is the relative size of the
clumps with respect to the radius. \textit{f$_c$} is the density enhancement
factor within the clump with respect to its environment. $\epsilon$ is the
flattening parameter of the spheroid shape of the clump (when $\epsilon=1$ it
is a sphere, $<1$ an oblate spheroid, and $>1$ a prolate spheroid). Based on
theoretical models \citep{1988PASA....7..434H,2005A&A...437..667D}, the
photosphere is formed at $\lesssim10^{15}$~cm, corresponding to
${\rm v_{ph}\lesssim500~km~s^{-1}}$, which is consistent with early-time
observations of SN~2024ggi \citep{2025A&A...699A..60E}, and is below $v_1$ for
the selected clumped models.
\end{minipage}

\end{table*}

In general, the starting conditions of our models begin with the overall properties of the supernova derived from observations. The time since the explosion is taken from the light curves, and the photosphere expansion velocity, $v_{photo}$, and effective temperature, $T_{eff}$, can be estimated from optical spectra  (see, e.g., \citealt{1974kirshner,2012takas,2014subhash,Hoogendam_etal_2024}). Additionally, when our spectra are analyzed, the distance to the object is taken at the currently accepted value, but we note that the distance uncertainties will propagate directly to CO mass uncertainties.

The temperature structure, $T(v)$, in the molecular-rich region with expansion velocities $v~ \epsilon~ [v_1,v_1 +\Delta v]$ is obtained under the assumption of radiative equilibrium being established by the cooling and heating of molecules (and dust). Finding $T(v)$ by the formal solution requires partial derivatives for all parameters. However, the seven-dimensional parameter space of MOFAT would make this method unstable. Instead, for any given set of parameters, the `optimal' temperature structure is calculated using a data-driven approach. Optimal means that $T(v)$ is found such that $\int \chi^2(\lambda) ~d\lambda$ between the observed and synthetic profiles is minimal. $T(v)$ at the inner boundary of the molecular region $v_1$ is determined by the observed photospheric spectrum. The temperature at the outer boundary is calculated under the assumption that the observed total flux in the molecular and dust bands is produced in the corresponding region for a given set of parameters.
Within the region, $T(v)$ is governed by radiative equilibrium, namely that the luminosity is balanced by the emissivity in each zone. However, we utilize $T(v)=T(v_1)+ C ~\Delta T(v)$  to force agreement with the flux observed at the outer boundary. This allows for a direct comparison between observed and synthetic line profiles and gauges the quality of the fits using $\chi^2$. $C$ is needed to compare the quality of profile fits, though $C$ for a given set may be unphysical. That said, this procedure allows us to find the seven-dimensional parameter set with the minimum $\chi^2$. Note that for the optimized solution of the full problem, $C$ approaches unity with a dispersion comparable to the noise in the observations.

The optimal set of free parameters related to the molecular features is found by a fixed point iteration in a seven-dimensional parameter space using a nested iteration scheme, with the order determined by the spectral tell-tails.

From the overtone, we find a temperature $T_1$ (defined at $v_1$), that best matches the observations because for 
optically thin features the inner high-density region dominates the emission. The overtone will also indirectly determine the total CO mass when we have a value for the slope, $n$, of the CO density distribution ($\rho\sim r^{-n}$). 
Throughout the envelope, the CO cooling is dominated by the fundamental band and drives \textbf{$\Delta T$} at the outer boundary.
For the inner loops, properties of the clumps are used.
We define our clumps as spheroids, where we can specify their morphology with three parameters: 1) $r/R$ is the relative size of the clumps with respect to any given radius in the envelope, 2) $f_c$ is the density enhancement factor within the clump with respect to its environment, and 3) $\epsilon$ is the flattening parameter of the spheroid (when $\epsilon =1$ it is a sphere, $< 1$ an oblate, and $ > 1$ a prolate). Oblate and prolate structures will have different surface-to-volume ratios and, thus, will have varying levels of absorption and emission. We will optimize our fits within both regions because they can have different physical interpretations. We note that the overtone can be fitted by a range of clumping parameters for each time, but the profiles of the fundamental bands cannot. Non-consistent parameters reveal themselves by failed fits of the fundamental (see below). MOFAT iterates all seven parameters until we converge on a set that best matches the observations\footnote{Solutions with reasonable parameters are clustered tightly.}.

\begin{figure*}
    \centering
    \includegraphics[width=0.9\linewidth]{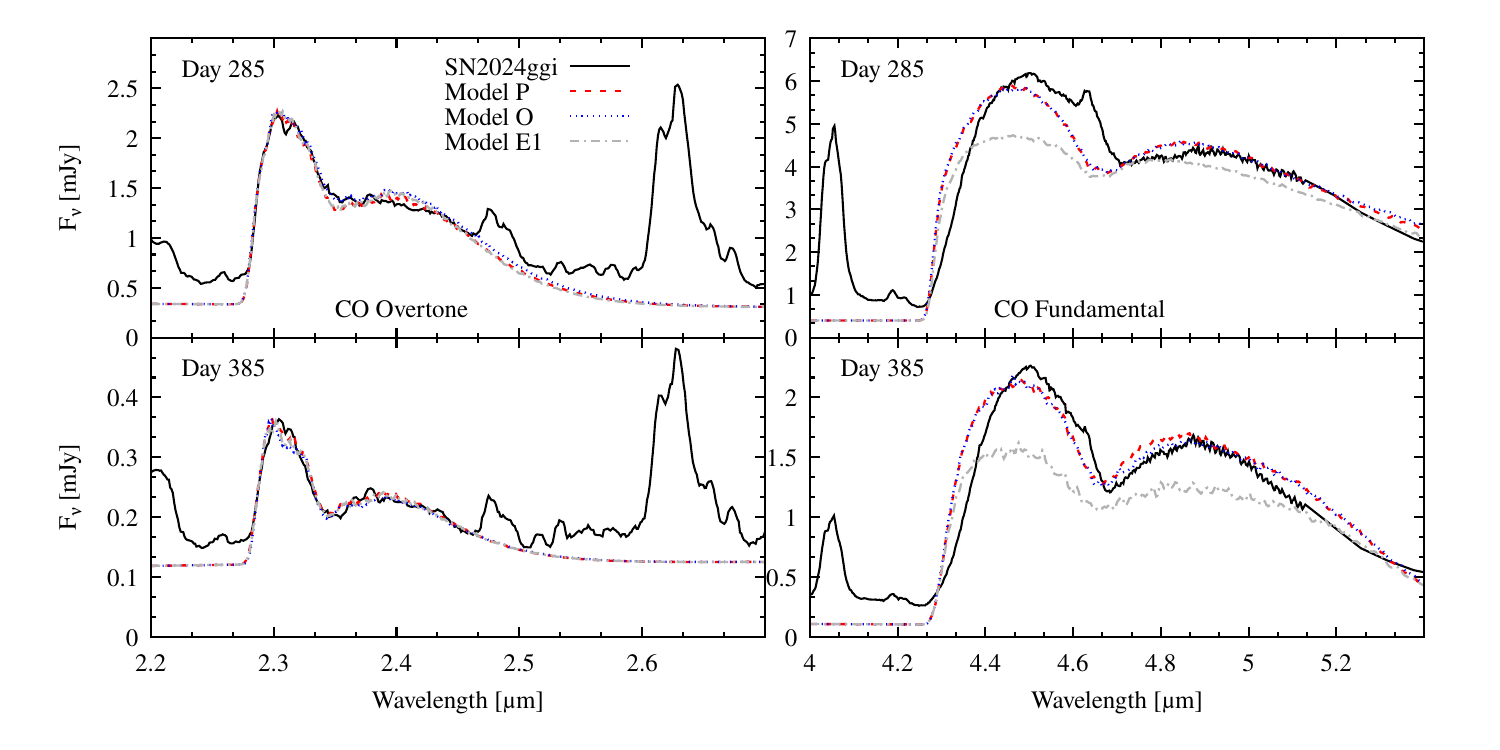}
    \caption{Comparing our prolate (P) and oblate (O) clumping models to the observations of \ggi. The upper panels show our results at 285 days, and the bottom panels show our results at 385 days. Our best fit models (P and O) are in dashed red and dotted blue, respectively, and we show the sensitivity of the parameters with the models defined with E1 (dot-dash grey). We vary the clumping size ($r/R$) at 285 days and the density enhancement ($f_c$) at 385 days. See Table \ref{tab:model_values} for the complete list of model parameters and Fig. \ref{fig:temp} for the temperature structure of the best fits. } 
    \label{fig:clumps}
\end{figure*}

\begin{figure*}[t]
    \centering
    \includegraphics[width=0.9\linewidth]{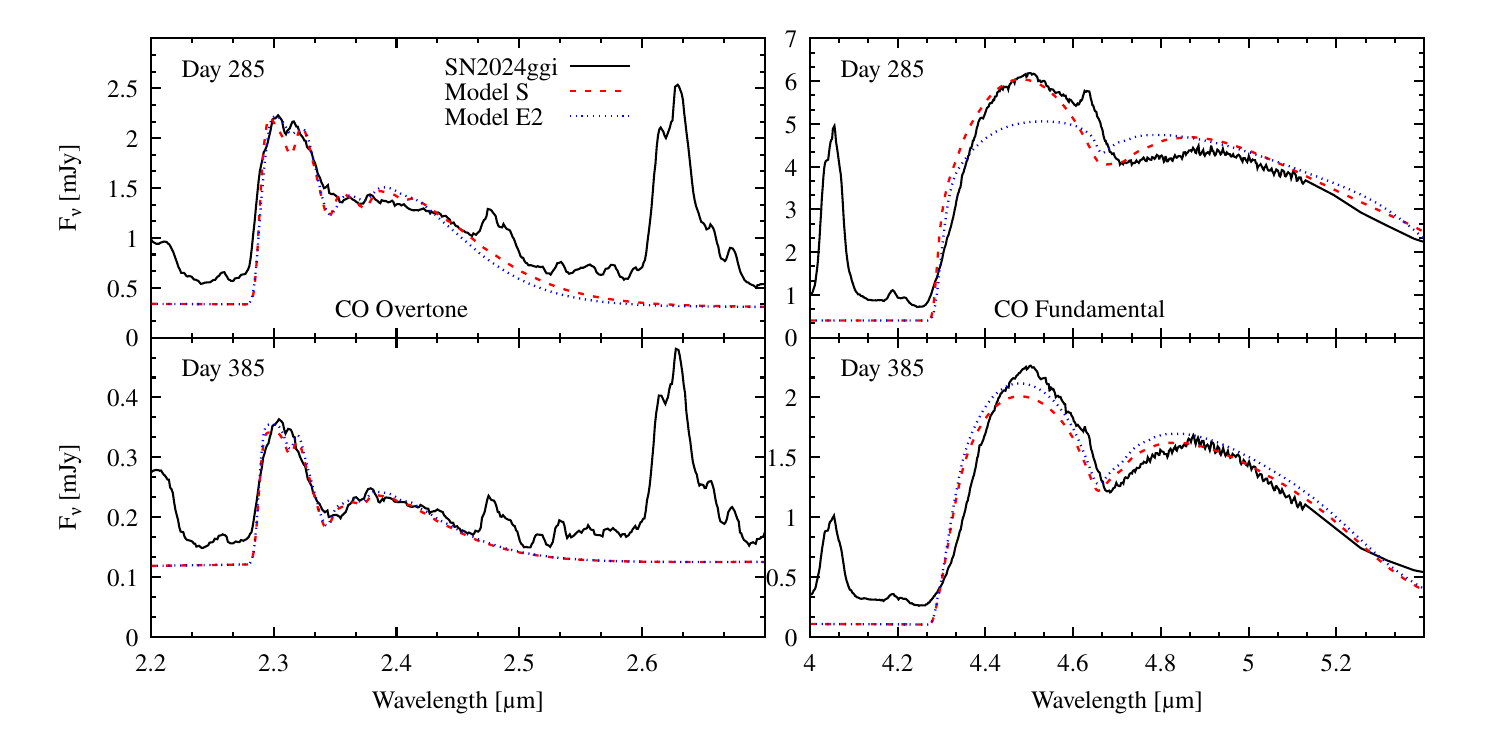}
    \caption{Same as Fig. \ref{fig:clumps} but for models without clumps. } 
    \label{fig:noclumps}
\end{figure*}
\begin{figure}
    \centering
    \includegraphics[width=0.85\linewidth]{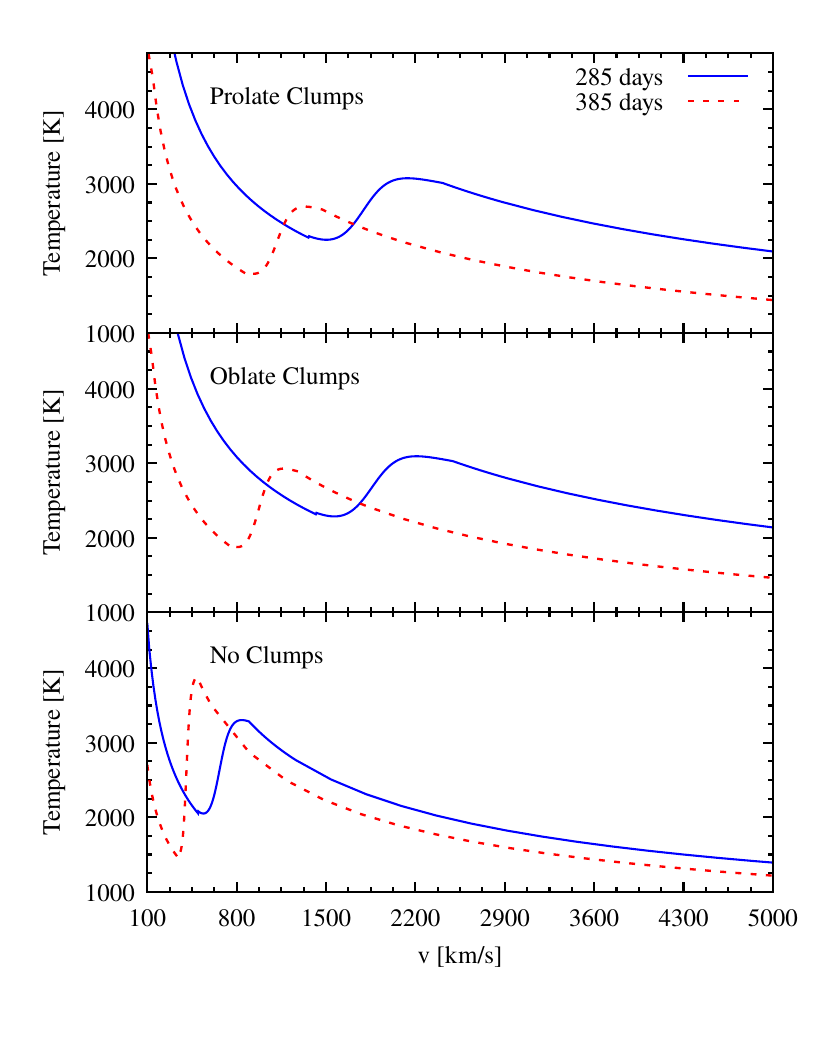}
    \caption{Converged temperature structures of gas with optimized parameters with and without clumps. In the upper panel, we show the prolate clumping model P. The middle panel shows the oblate clumping model O. The lower panel shows the non-clumping model S. See Table \ref{tab:model_values} for the model parameters and figures \ref{fig:clumps} and \ref{fig:noclumps} for their fits.}
    \label{fig:temp}
\end{figure}
\begin{figure*}
    \centering
    \includegraphics[width=0.95\linewidth]{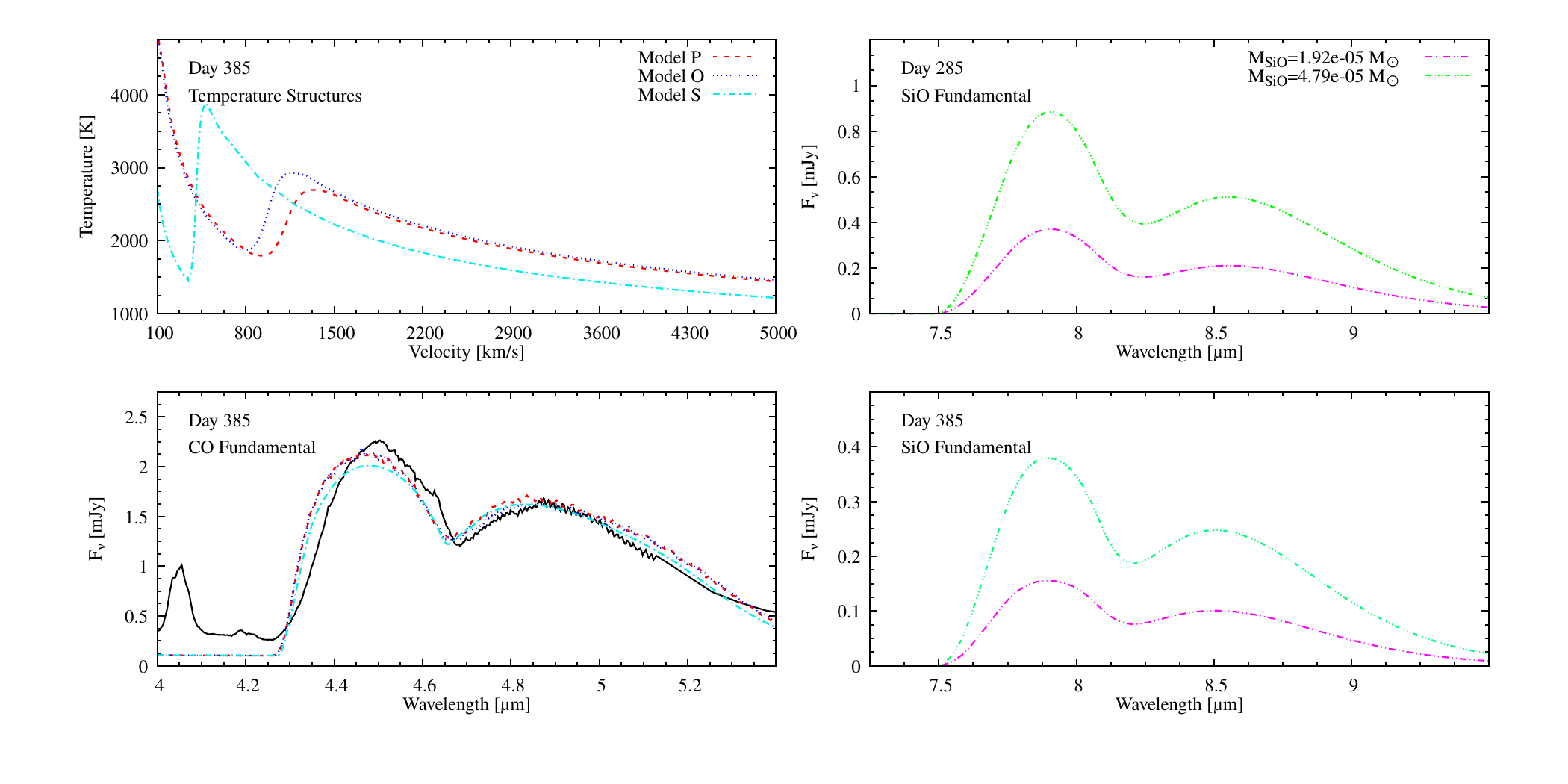}
    \caption{We compare the temperature profiles of the best fit models at +385 d in the upper left quadrant. In the lower left panel we compare the fundamental CO band of the models with and without clumps at +385 d. Without clumps, the ratio between the first and second vibrational modes of the fundamental band are off by $\approx 15 \%$. 
    In addition (upper and lower right quadrants), we give synthetic SiO fundamental profiles using the temperature structures of Model P (Fig. \ref{fig:temp}) at +285 d and +385 d, respectively. We use equilibrium abundances 
    based on the $15 M_\odot$ stellar model with solar metallicity \citep{2003chieffi} and find that 
    most of the Si exposed is in atomic Si and only some $6.3 \times 10^{-4} M_\odot$ and $1.5 \times10^{-3} M_\odot$ are in SiO at +285 d and +385 d, respectively. For the synthetic spectra we do not include clumping and use: $n=1.1$, $v_1=900 ~\mathrm{km/s}$, and $\Delta v=900~\mathrm{km/s}$ at both epochs. 
    We did not demonstrate parameter differences because Si- and C-rich layers can be expected to have different clump structures, and we assumed equilibrium for the SiO estimate.} 
    \label{fig:noclumpsvsclumps}
\end{figure*}

\subsection{Analysis of the Molecular Features}\label{sec:AnaSpec}

In Figures \ref{fig:clumps} (models P and O) and \ref{fig:noclumps} (model S), we show our best fits of both the fundamental and overtone CO bands for SN~2024ggi for prolate, oblate, and non-clumping models, respectively, at +285 and +385 days. E1 and E2 are given as examples for clumped and non-clumped models where their parameters fail to fit the profile despite temperature structure convergence based on a consistent molecular cooling function. The model parameters are given in Table \ref{tab:model_values}.

Several properties are common between all models shown: The CO-mass decreases by a factor of $\approx 8$ between observations, the location of the CO-rich region in velocity space is very similar at either time, the minimum velocity, $v_1$, decreases with time, and the reference temperature, $T_1$, decreases with time, with overall differences between the temperature profiles shown in Fig. \ref{fig:temp}. The temperature minimum is mainly caused by the cooling of the optically thick fundamental band.  
Further out, $T$ rises due to the fundamental band becoming optically thin from the increasing blueshift and geometrical dilution allowing for more exposure to intense radiation from the inner ejecta. This increases the heating of the inner CO layers and the dissociation of CO 
 and provides a possible explanation for the CO destruction.
Full radiation-hydro models that included CO formation show that most of the CO is formed close to the photosphere (at temperatures around $T_1 \approx 5000$ K) with CO freely expanding with ejecta; \citealt{hoeflich_1988,Gerard_etal_2000,Rho_2021}). Together with the reference temperature being much smaller than the formation temperature, the decreasing mass of CO, and the inner velocity only slightly decreasing, suggest that the CO destruction rate outpaces the formation rate, and that the photosphere has receded to the carbon-poor, Si-rich layers of the explosion. 

We find good overall agreement in the fits of clumping and non-clumping models at 285 days (Fig. \ref{fig:clumps} for P and O, and \ref{fig:noclumps} for S, respectively), but
only P and O allow for a good reproduction of the fundamental band at the later date. At each epoch, the CO overtone can be reproduced in all examples, demonstrating the need for the MIR observations to separate optically thin, one-temperature models to those that include optical depth effects. One exception is the small dip at 2.31 ${\rm \mu m}$ in the vibrational modes of the overtone. This dip is present in our simulations, and is
produced by either self-absorption in solutions that are barely optically thick, or by varying degrees of temperature decoupling between the vibrational and rotational bands. Similar structures have been seen in SN~1987A \citep{Liu_Dalgarno_1992}.
The feature at $4.65 {\rm \mu m}$ is the Pf$_{\beta}$ and is not included in our fits (see Fig. \ref{fig:data}).

Note that at +285 d the minimum temperature by CO cooling is slightly above $2000$~K in a narrow range for our clumping models (see Fig. \ref{fig:temp}). This temperature is approaching the onset of SiO formation (see Fig. 18 of \citealt{Rho_2021}). By +385 d the minimum temperature is below 2000 K, but at this T, only little Si will be bound in SiO which would be insufficient to effectively cool. If we assume equilibrium abundance based on the $15 M_\odot$ stellar model with solar metallicity from \cite{2003chieffi} and take the temperature profiles of model P, we would expect to find some $4.8 \times 10^{-5} M_\odot$ and $1.9 \times10^{-5} M_\odot$ of SiO at +285 d and +385 d, respectively. 
In Fig. \ref{fig:noclumpsvsclumps}, we give synthetic spectra of the SiO fundamental band using these mass estimates in a narrowed molecular region.

The similarity between the clumped and non-clumped models at the early epoch can be understood as an optical depth degeneracy. In the non-clumped model, the CO-forming material is confined to a narrower velocity interval, with $\Delta v\approx764~\mathrm{km~s^{-1}}$ compared with $\Delta v\approx1500~\mathrm{km~s^{-1}}$ in the clumped models. Concentrating the CO within this smaller emitting volume provides sufficient optical depth to reproduce the absolute flux of the fundamental band, while the high temperature (lower opacities) reproduces the relative strengths of its first two vibrational components. Conversely, the clumped models distribute the CO over a broader velocity range, while the local density enhancement, $f_c\approx5.5$, increases the optical depth and self-absorption within the clumps. This suppresses the fundamental-band flux that would otherwise be produced by the larger emitting volume. Thus, the two geometries reproduce similar early-time spectra through different combinations of emitting volume and optical depths.

However, at 385 days, we find consistent agreement between the clumping models and the observations of \ggi, but found poor agreement for our converged non-clumping model (see Fig. \ref{fig:noclumpsvsclumps} for a direct comparison). 
In detail, Model S shows shortcomings in the fundamental band, namely the peak-flux ratio between the first and second vibrational components is $\sim8\%$ smaller than observed, compared with $\sim3\%$ for the clumped models.
 More precisely, in smooth distributions the ratio between the opacities of the first and second mode and the radii at which they become optically thin make their flux ratio stable. In clumpy media, the emission is formed over a wide range ('picket fence') and, thus,
 it depends sensitively on the morphology and over-density of the clumps providing a sensitive diagnostic.
 Because models with clumps converge early on and are needed at later time, we always include clumps as free parameters and discuss their evolution.

 For both O and P models, there is a tendency for the clumps to become smaller and rounder with increasing density contrast $f_c$.
 With time, CO in the outer layers of the clumps becomes dissociated by supernova radiation (see above), leading to both rounder and smaller clump sizes. The higher $f_c$ is suggested
by the pressure equilibrium between the heated outer and the cool central layers. 
 The two dates are insufficient to distinguish oblate and prolate structures.

 We have demonstrated the utility of the inverse-data-driven method of analysis. For future observations, low-latency observations are needed to evaluate the process of CO formation right after the onset, namely, the end of the plateau phase, the subsequent formation of SiO, and at later times to answer whether molecule formation is the trigger of dust formation in SNe.

\section{Conclusions}  \label{sec:Conclusions}

We present NIRSpec \jwst\ data of \ggi\ during the nebular phase at +285.64~d and +385.55~d post-explosion, obtained as part of our Director's Discretionary Program 6717 \citep{Ashall6716A}. The main results of the paper are:

\renewcommand{\labelenumi}{(\arabic{enumi})}
\begin{enumerate}

\item
The spectra are dominated by atomic emission lines from H, Ca, Ar, C, Mg, Ni, Co, and Fe. These lines exhibit asymmetry, suggesting an asymmetric chemical distribution in \ggi\ (Fig. \ref{fig:speccomp} and Sect. \ref{sec:Spec}).

\item
The spectra also reveal clear signatures of CO through the emission of flux from the first overtone and fundamental bands (Sect. \ref{sec:Spec}).

\item 
We presented a new, data-driven approach that directly targets the properties of the ejecta with molecules \citep{Mera_2025_theory} using non-LTE transport simulations to calculate the temperature structure in the molecular-rich region of the ejecta. We discussed the sensitivity of this method to detect and characterize inhomogeneities (Sect.\ref{sec:Models}), which allows us to
decipher the complex 3D structure and reproduce detailed high-fidelity profiles.
This approach fundamentally differs and overcomes the two classical approaches which are I) assuming one-zone models with a given temperature in the optically thin case, which is not valid for the fundamental molecular band, and II) the forward approach that calculates the full explosion in spherical geometry. Though justified for the formation of CO, it fails to capture the complexity of the physics seen at later times  (Sect. \ref{sec:AnaSpec}).

\item 
The CO first overtone is mostly optically thin at both dates. MIR spectra (i.e. \jwst) are crucial 
for deciphering the structure and provide stable mass-estimates for molecules. However, a small inversion on top of the overtone band is reproduced as an optical depth combined with a non-LTE effect in the CO-cooling region. (Figs. \ref{fig:clumps} and \ref{fig:noclumps}). 

\item
At +285~d, the model parameters indicate large-scale, slightly clumped CO distributions. Both
models with and without small-scale structure give reasonable fits to the data, with higher quality requiring clumps.
The spectrum at +385~d clearly demonstrates the importance of clumps and the need for spectral series (Sect. \ref{sec:AnaSpec} and Fig. \ref{fig:noclumpsvsclumps}).

\item
The majority of CO formation is before +285 d and the subsequent phase is characterized by
CO-destruction from ($M_{CO}(+285~d) \approx 8\times 10^{-4} M_\odot$  to $ M_{CO}(+385~d) \approx 1.0 \times 10^{-4} M_\odot$). In our models, this 
can be attributed to the rising temperature caused by energy input from inner layers because
doppler shift and geometrical dilution exposes CO to harder radiation and heating (Fig. \ref{fig:temp} and Sect. \ref{sec:AnaSpec}).

\item  The clumps in our models are optically thick in the fundamental band, leading to smaller and rounder clumps as CO is destroyed, where the pressure equilibrium between the hot outer and cool inner layers increases the density enhancement factor ($f_c$) (Tab. \ref{tab:model_values}).

\item 
The inner edge of the CO-rich region is well above the photosphere at both epochs, suggesting
that the photosphere has entered the region of advanced burning stellar products. From model based estimates we might expect to find $M_{SiO}\approx4.8 \times 10^{-5} M_\odot$ and $1.9 \times10^{-5} M_\odot$ at +285 d and +385 d, respectively (Fig. \ref{fig:noclumpsvsclumps}). Although masses of this level would be insufficient to reasonably cool the ejecta, we suggest that more SiO will be formed at later times and may trigger additional cooling leading to dust formation.

\end{enumerate}

Finally, we also have to mention the shortcomings. Early time observational coverage after the plateau phase would be needed to study the formation process and to link
the CO-rich layers to the explosion physics, e.g., by the observation of ionized
molecules \citep{Spyromilio_1988,Rho_2021}.
Later times with good time coverage are needed to probe the phase of strong SiO formation with additional cooling. This is needed to establish molecular formation as a trigger for dust formation.
From theory, an extended study of the sensitivity on the various parameters is underway to develop further diagnostics, e.g., linking the SiO bands and their appearance to the progenitor and explosion physics. Taking the new constraints from above, it is tempting to employ complex multi-D simulations which use a forward approach, but, SNe~II are a very diverse group of objects, so, a large grid of models and many well-sampled SNe~II would be needed to understand the growth and cooling of small scale 3D structures.

Overall, we have demonstrated that \jwst\ spectral observations of SNe~II can be used to infer the three-dimensional geometry of the molecule-forming regions. In the case of \ggi, clumping is required to simultaneously reproduce both the CO first overtone and fundamental band. Such clumping may be a common feature in all SN~II explosions. Looking ahead, time-series spectral observations beginning shortly after a SN~II exits the plateau phase would allow for direct tracking of both the growth of CO mass and the evolution of clumping within the ejecta.

\begin{acknowledgments}
We thank the director of the Space Science Institute for granting time for these observations. 
T.M., K.M., E.B., C.A., J.D., M.S., and  P.H. acknowledge support from NASA grants JWST-GO-02114,
JWST-GO-02122, JWST-GO-04522, JWST-GO-04217, JWST-GO-04436,
JWST-GO-03726, JWST-GO-05057, JWST-GO-05290, JWST-GO-06023,
JWST-GO-06677, JWST-GO-06213, JWST-GO-06583, JWST-GO-0923 Support for
programs \#2114, \#2122, \#3726, \#4217, \#4436, \#4522,  \#5057,
\#6023, \#6213, \#6583, and \#6677
were provided by NASA through a grant from the Space Telescope Science
Institute, which is operated by the Association of Universities for Research in
Astronomy, Inc., under NASA contract NAS 5-03127.ACKNOWLEDGMENTS

P.H. an T.M. acknowledge the support by the National Science Foundation NSF awards AST-1715133 and  AST- AST-230639 for enabling the development of the methods and computational codes
and, in parts, covering the graduate and postdoc salaries.

L.G. acknowledges financial support from AGAUR, CSIC, MCIN and AEI 10.13039/501100011033 under projects PID2023-151307NB-I00, PIE 20215AT016, CEX2020-001058-M, ILINK23001, COOPB2304, and 2021-SGR-01270.

JTH acknowledges support from NASA through the NASA Hubble Fellowship grant HST-HF2-51577.001-A, awarded by STScI. STScI is operated by the Association of Universities for Research in Astronomy, Incorporated, under NASA contract NAS5-26555

W.B.H. acknowledges support from the NSF Graduate Research Fellowship Program under Grant No. 2236415. 

\end{acknowledgments}

\noindent
{\sl Facilities:}
The observations were obtained with the \jwst.
The simulations were performed on the parallel-cluster of Florida State University.

\bigskip
\noindent
{\sl Software:}
MOFAT, part of a PhD thesis, and selective HYDRA modules used in this work are available by request after the PhD thesis is finalized. 
\bigskip
\noindent

\noindent
{\sl Data Availability Statement:} The \jwst data for \ggi\ were obtained from the Mikulski Archive for Space Telescopes. These observations can be accessed via \dataset[doi:10.17909/3v0v-0a76]{https://doi.org/10.17909/3v0v-0a76}.
All other data is available on request.
\bibliographystyle{aasjournal}

\input{main.bbl}
\end{document}

%% file: authors.tex
\newcommand{\PSI}{\affiliation{Planetary Science Institute, 1700 East Fort
  Lowell Road, Suite 106,Tucson, AZ 85719-2395 USA}}
\newcommand{\HS}{\affiliation{Hamburger Sternwarte, Gojenbergsweg 112, 21029 Hamburg, Germany}}
\newcommand{\IFA}{\affiliation{Institute for Astronomy, University of Hawai’i at Manoa, 2680 Woodlawn Dr., Hawai’i, HI 96822, USA}}
\newcommand{\VT}{\affiliation{Department of Physics, Virginia Tech,
    850 West Campus  Drive, Blacksburg VA, 24061, USA}}
\newcommand{\GRFP}{\altaffiliation{National Science Foundation Graduate Research Fellow}}

\newcommand{\FINESST}{\altaffiliation{NASA FINESST Future Investigator}}
\newcommand{\NHFPE}{\altaffiliation{NHFP Einstein Fellow}}
\newcommand{\UIUC}{\affiliation{Department of Astronomy, University of Illinois Urbana-Champaign, 1002 West Green Street, Urbana, IL 61801, USA}}
\newcommand{\NSFSIMS}{\affiliation{NSF-Simons AI Institute for the Sky (SkAI), 172 E. Chestnut St., Chicago, IL 60611, USA}}

\newcommand{\STSci}{\affiliation{Space Telescope Science Institute, 3700 San Martin Drive, Baltimore, MD 21218-2410, USA}}
\newcommand{\FSU}{\affiliation{Department of Physics, Florida State
    University, Tallahassee, FL 32306, USA}}
\newcommand{\Carnegie}{\affiliation{Observatories of the Carnegie
    Institution for Science, 813 Santa Barbara St., Pasadena, CA 91101, USA}}
\newcommand{\MSU}{\affiliation{Department of Physics \& Astronomy,
    Michigan State University, East Lansing, MI, USA}}
\newcommand{\TAMU}{\affiliation{George P. and Cynthia Woods Mitchell
    Institute for Fundamental Physics and Astronomy,
    Department of Physics and Astronomy, Texas 
             A\&M University, College Station, TX 77843, USA}}
\newcommand{\IALP}{\affiliation{Instituto de Astrof\'isica de La Plata
    (IALP), CONICET, Paseo del Bosque S/N, B1900FWA La Plata, Argentina}}
\newcommand{\LaPlata}{\affiliation{Facultad de Ciencias Astron\'omicas
    y Geof\'isicas Universidad Nacional de La Plata, Paseo del Bosque,
    B1900FWA, La Plata, Argentina}}
\newcommand{\WPI}{\affiliation{Kavli Institute for the Physics and
    Mathematics of the Universe (WPI), The University of Tokyo,
    Kashiwa, 277-8583 Chiba, Japan}} 

\newcommand{\ICE}{\affiliation{Institute of Space Sciences (ICE,
    CSIC), Campus UAB, Carrer de Can Magrans, s/n, E-08193 Barcelona, Spain}}

\newcommand{\IEEC}{\affiliation{Institut d’Estudis Espacials de
    Catalunya (IEEC), E-08034  Barcelona, Spain}} 

\newcommand{\LCO}{\affiliation{Las Campanas Observatory, Carnegie
    Observatories, Casilla 601, La Serena, Chile}} 

\newcommand{\Aarhus}{\affiliation{Department of Physics and Astronomy,
    Aarhus University, Ny  Munkegade 120, DK-8000 Aarhus C, Denmark.}} 

\newcommand{\OU}{\affiliation{Homer L.~Dodge Department of Physics and
  Astronomy, University of Oklahoma, 440 W. Brooks, Rm 100, Norman, OK
  73019-2061}}  

\newcommand{\UCSC}{\affiliation{Department of Astronomy and Astrophysics,
  University of California, Santa Cruz, CA 95064, USA}} 
\newcommand{\Melbourne}{\affiliation{School of Physics, The University of
  Melbourne, VIC 3010, Australia}}

\newcommand{\LPNHE}{\affiliation{LPNHE, (CNRS/IN2P3, Sorbonne
  Universit\'e, Universit\'e Paris Cit\'e), Laboratoire de Physique
  Nucl\'eaire et de Hautes \'Energies, 75005, Paris, France}}

\newcommand{\Princeton}{\affiliation{Princeton University, 4 Ivy Lane,
    Princeton, NJ 08544, USA}}

\newcommand{\Berkeley}{\affiliation{Department of Astronomy,
    University of California, Berkeley, CA 94720-3411, USA}}

\newcommand{\Tsinghua}{\affiliation{Physics Department, Tsinghua
    University, Beijing, 100084, China}}

\newcommand{\Thailand}{\affiliation{National Astronomical Research
    Institute of Thailand, 260 Moo 4, Donkaew, Maerim, Chiang Mai,
    50180, Thailand}}

\newcommand{\UVA}{\affiliation{Department of Astronomy, University of
    Virginia, 530 McCormick Rd, Charlottesville, VA 22904, USA}}

\newcommand{\LJMU}{\affiliation{Astrophysics Research Institute,
    Liverpool John Moores University, 146 Brownlow Hill, Liverpool L3
    5RF, UK}}

\newcommand{\MPIA}{\affiliation{Max-Planck-Institut f\"ur Astrophysik,
    Karl-Schwarzschild Stra{\ss}e 1, 85748 Garching, Germany}}

\newcommand{\JHU}{\affiliation{Physics and Astronomy Department,
    Johns Hopkins University, Baltimore, MD 21218, USA}}

\newcommand{\OSU}{\affiliation{Department of Astronomy, The Ohio State
    University, Columbus, OH, USA}}

\newcommand{\CCAP}{\affiliation{Center for Cosmology and Astroparticle
    Physics, The Ohio State University, Columbus, OH, USA}}

\newcommand{\MIT}{\affiliation{Department of Physics and Kavli Institute for Astrophysics and Space Research, Massachusetts Institute of Technology, 77 Massachusetts Avenue, Cambridge, MA 02139, USA}}

\newcommand{\nextinstitute}{\affiliation{Put the institute of the new author here}}

\author[0000-0001-5888-2542]{T.~Mera}
\email{tycomera@gmail.com}
\FSU

\author[0000-0002-5221-7557]{C. Ashall}
\email{cashall@hawaii.edu}
\IFA

\author[0000-0002-4338-6586]{P.~Hoeflich}
\email{phoeflich77@gmail.com}
\FSU

\author[0000-0001-7186-105X]{K. Medler}
\email{kyle.medler@sky.com}
\IFA

\author[0000-0002-9301-5302]{M.~Shahbandeh}
\email{mshahbandeh@stsci.edu}
\STSci

\author[0000-0003-4625-6629]{C.~R.~Burns}
\email{cburns@carnegiescience.edu}
\Carnegie

\author[0000-0001-5393-1608]{E.~Baron}
\email{ebaron@psi.edu}
\PSI
\HS

\author[0000-0002-7566-6080]{J. M. DerKacy}
\email{jderkacy@stsci.edu}
\STSci

\author[0000-0002-4338-6586]{N.~Morrell}
\email{nmorrell@carnegiescience.edu}
\LCO

\author[0000-0002-3900-1452]{J. Lu}
\email{lujingeve158@gmail.com}
\MSU

\author[0000-0001-9668-2920]{J.~T.~Hinkle}
\NHFPE
\UIUC
\NSFSIMS
\IFA
\email{jhinkle6@hawaii.edu}


\author[[0000-0001-6876-8284]{P. A. Mazzali}
\email{P.Mazzali@ljmu.ac.uk}
\LJMU
\MPIA

\author[0009-0001-9148-8421]{E.~Fereidouni}
\email{ef22g@fsu.edu}
\FSU

\author[0000-0002-7305-8321]{C.~M.~Pfeffer}
\GRFP
\IFA
\email{cpfeffer@hawaii.edu}

\author[0000-0001-6107-0887]{S. Shiber}
\FSU
\email{sshiber1@lsu.edu}

\author[0000-0001-7380-3144]{T. Temim}
\email{temim@astro.princeton.edu}
\Princeton


\author[0000-0002-1296-6887]{L. Galbany}
\email{lluisgalbany@gmail.com}
\ICE
\IEEC
  
\author[0000-0003-4263-2228]{D.~A.~Coulter}
\email{dcoulter@stsci.edu}
\JHU
\STSci


\author[0009-0000-6303-4169]{L. Ferrari}
\email{luciaferrari4@gmail.com}
\IALP
\LaPlata





\author[0000-0003-3953-9532]{W.~B.~Hoogendam}
\GRFP
\IFA
\email{willemh@hawaii.edu}

\author[0000-0003-1039-2928]{E.~Y.~Hsiao}
\email{yichi.hsiao@gmail.com}
\FSU






\author[0000-0003-2734-0796]{M.~M.~Phillips}
\email{mmp@lco.cl}
\LCO

 

\author[0000-0003-4631-1149]{B.~J.~Shappee}
\email{shappee@hawaii.edu}
\IFA








